%% file: manuscript.tex
\documentclass{egpubl}
\usepackage{eurovis2024}

\SpecialIssuePaper         

\CGFccby

\usepackage[T1]{fontenc}
\usepackage{dfadobe}  

\usepackage{cite}  
\BibtexOrBiblatex
\electronicVersion
\PrintedOrElectronic
\ifpdf \usepackage[pdftex]{graphicx} \pdfcompresslevel=9
\else \usepackage[dvips]{graphicx} \fi

\usepackage{egweblnk}

\newcommand{\ourTerm}{GerontoVis\xspace}
\newcommand{\numPapers}{36\xspace}
\newcommand{\numVenues}{29\xspace}

\usepackage[LGR,T1]{fontenc}
\usepackage[utf8]{inputenc}   
\newcommand{\textgreek}[1]{\begingroup\fontencoding{LGR}\selectfont#1\endgroup}

\usepackage{xspace}
\usepackage{subcaption}
\usepackage[nameinlink]{cleveref}
\usepackage{ragged2e}
\usepackage{breakurl}

\usepackage{soul,xcolor}
\setstcolor{red}

\usepackage[normalem]{ulem}

\title[\ourTerm: Data Visualization at the Confluence of Aging]%
      {\ourTerm: Data Visualization at the Confluence of Aging}

\author[Z. While, R.J. Crouser, \& A. Sarvghad]
{\parbox{\textwidth}{\centering Zack While$^{1}$,\thanks{Corresponding author: zwhile@cs.umass.edu}
\orcid{0000-0002-9114-3984}
R. Jordan Crouser$^{2}$,\orcid{0000-0001-9936-0791} 
        and Ali Sarvghad$^{1}$
        \orcid{0000-0003-3718-7043} 
        }
        \\
{\parbox{\textwidth}{\centering $^1$Manning College of Information and Computer Sciences, University of Massachusetts, Amherst, USA\\
         $^2$Department of Computer Science, Smith College, Northampton, USA
       }
}
} 

\begin{document}

\maketitle
\begin{abstract}
   Despite the explosive growth of the aging population worldwide, older adults have been largely overlooked by visualization research. 
   This paper is a critical reflection on the underrepresentation of older adults in visualization research. We discuss why investigating visualization at the intersection of aging matters, why older adults may have been omitted from sample populations in visualization research, how aging may affect visualization use, and how this differs from traditional accessibility research. To encourage further discussion and novel scholarship in this area, we introduce \textbf{\ourTerm}, a term which encapsulates \emph{research and practice of data visualization design that primarily focuses on older adults.} By introducing this new subfield of visualization research, we hope to shine a spotlight on this growing user population and stimulate innovation toward the development of aging-aware visualization tools. We offer a birds-eye view of the \ourTerm landscape, explore some of its unique challenges, and identify promising areas for future research.


\end{abstract}  

\input{1-introduction}

\input{2-why-has-vis-research-overlooked-older-adults}

\input{3-gerontovis-vs-accessibility}

\input{4-age-related-physiological-changes}

\input{5-state-of-the-practice}

\input{6-challenges-and-opportunities}

\input{7-conclusion}

\section*{Acknowledgements}
    The authors wish to thank Dr. Thalia Pandiri, Professor of Classical Languages \& Literatures at Smith College, for her counsel regarding the conjugation of the Greek root \textgreek{γέρων}.
\clearpage

\bibliographystyle{eg-alpha-doi} 
\bibliography{bibliography}


\end{document}

%% file: 1-introduction.tex
\section{Introduction}
Advances in healthcare, nutrition, and living conditions have led to dramatic increases in human life expectancy.
The World Health Organization predicts that the global population of older adults will increase from $\approx$1 billion in 2020 to $\approx$2.1 billion by 2050~\cite{Ageingan92:online}.
The number of older adults will match the number of children age 12 and younger by 2050~\cite{unWorldPopulation}, 
and the U.S. Bureau of Labor estimates that by 2030, eight percent of the workforce in the United States will be aged 65 or older~\cite{Numberof42:online}. This demographic shift has notably increased attention to designing and building data-driven technological interventions that empower older adults in various aspects of their social and personal lives~\cite{mannheim2019inclusion}. 

While explorations into various perceptual, cognitive, and psychosocial aspects of data visualization have resulted in a wealth of empirical knowledge and design guidelines~\cite{meloncon2017data,gubala2022data,franconeri2021}, 
this research has notably overlooked older adults. 
Studies that center this demographic as primary end users are remarkably sparse~\cite {backonja2016visualization,le-2016-eval}, despite the prevalent use of data visualizations to assist older adults in monitoring and managing various aspects of their health~\cite{le-2016-eval}, physical activities  (e.g., \cite{fanning2018mobile,gualtieri2016can,kimura2019modifiable,tapia2016designing}), and making data-informed decisions (e.g., \cite{garcia2012using, galesic2009using, ruiz2013communicating,price2016effects}). In a recent survey of mobile health technology for older adults, Cajamarca et al.~\cite{cajamarca2020} found that data visualizations were part of 92.6\% of interventions; however, only 23.5\% of studies assessed the suitability of various visualization features for this population. 

Of course, there is no universal agreement on what constitutes \emph{old}. The United Nations (UN) uses the term ``older persons'' to refer to those aged 60 years and over~\cite{UNHCR} in its policies and programs related to aging and older persons. The World Health Organization uses the term ``older adults'' to refer to individuals who are aged 65 years and over.
While many US-based studies consider ``older adults'' to be individuals 65 years of age and above~\cite{CDC-older}, 
beliefs about what constitutes ``old age'' seem to be fluid and appear to change as people age~\cite{cameron1969age,drevenstedt1976perceptions,hori1994beginning}. 
Rather than attempting to establish a rigid threshold regarding the chronological age at which someone should be considered ``old'', in this work we will use the term ``older adults'' to refer to all individuals who have experienced perceptual, cognitive, and/or physical changes due to the natural physiological process of aging.

In this \textit{position paper and call to action}, we introduce the term \textbf{\ourTerm} to call attention to the critical knowledge gap at the intersection of aging and visualization research, and to encourage future investigation in this area.
Borrowing the Greek root \textgreek{γέρων} meaning ``elder'' (anglicized genitive \textit{gerontos}), we formally define \ourTerm as \textit{``a sub-field of visualization, encompassing research and practice of data visualization that primarily focuses on older adults''.} \ourTerm is concerned with \textbf{identifying barriers that may prevent older adults from accessing or effectively using data visualizations, and developing best practices for design}. This paper seeks to identify ways in which a shift toward explicit inclusion of older adults could benefit VIS research as a whole, as well as explore the intersection of \ourTerm with accessible design more broadly.

\begin{figure}[t!]
    \ \ \ \ \includegraphics[width=0.85\columnwidth]{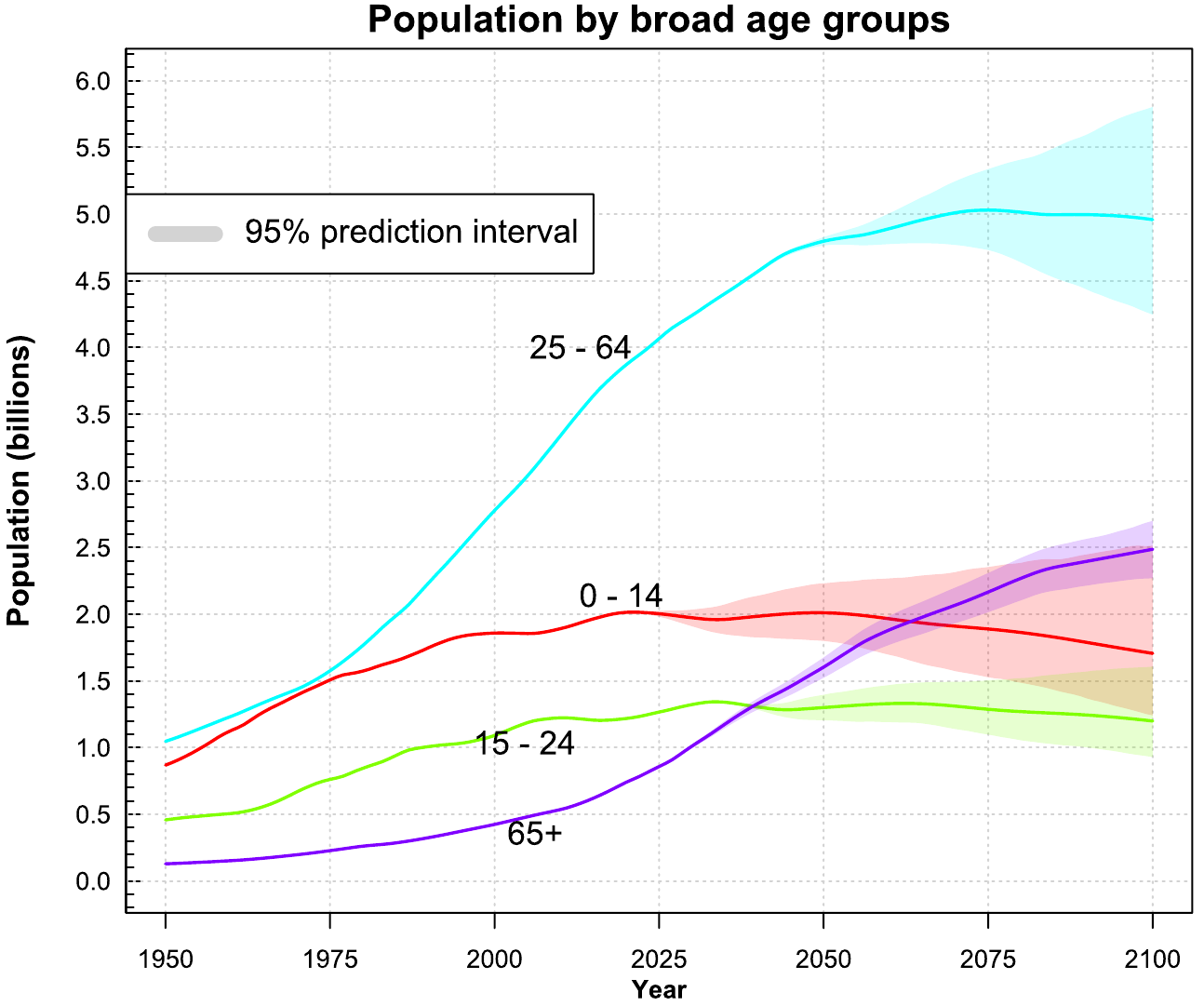}
    \caption{The global population of older adults is expected to surpass the number of children (age 0-14) between 2050 and 2075. Source:~\cite{WorldPop57:online}, CC BY 3.0 IGO.}
    \label{fig:global_population_trends}
\end{figure}

The primary contribution of this work lies in the formal establishment of \ourTerm, a novel subfield of visualization research specifically centered on the needs of older adults. We critically reflect on some reasons for older adults' historical exclusion from previous visualization research, and argue why it is critical to distinguish \ourTerm from broader accessibility research.  
Finally, we contribute an initial set of research directions, offering a starting point for those interested in \ourTerm research.
The remainder of this paper is organized as follows. In \Cref{why-underrep}, we reflect on some of the reasons that may contribute to the underrepresentation of older adults in visualization research. \Cref{accessibility} describes the key distinctions between \ourTerm and research on accessible visualization. In \Cref{age-related-changes}, we highlight how a number of age-related physiological changes may interact with older adults' ability to use visualizations. This is followed by a concise review of existing visualization research that focuses on older adults in \Cref{survey}. It is important to clarify that our objective was \textit{not} to conduct a systematic review of existing literature but rather to present a cross-section of examples to provoke constructive discourse on the imbalance between the growing size of the older adult population and their representation as participant-stakeholders in visualization research. Finally, \Cref{challenges-and-opportunities} provides a discussion of \ourTerm research opportunities and challenges.

%% file: 2-why-has-vis-research-overlooked-older-adults.tex
\section{Why has VIS Research Overlooked Older Adults?}
\label{why-underrep}

 Answering this question requires close investigation and in-depth examination of the visualization research community's agenda, practices, and values. However, the underrepresentation of older adults is not unique to visualization research and exists in many scientific disciplines~\cite{nelson2004ageism}. Hence, drawing on prior work regarding this issue in science (e.g.,~\cite{mody2008recruitment,forsat2020recruitment,bowling20195ts}), we offer a few thoughts on how and why the underrepresentation of older adults can manifest in visualization research.

\subsection{Cascading Exclusion Criteria}
How research studies are designed can inadvertently contribute to the non-random exclusion of certain groups. For example, it is common for studies in VIS (and HCI more broadly) to impose vision-related restrictions on study populations, e.g., requiring 20/20 or corrected-to-20/20 vision as explicit inclusion criteria. This criterion, most likely intended to isolate the effects of differences in visual acuity, has been established as a sort of "cultural norm" even in studies where such factors have not been observed or hypothesized to bias the result.
However, because many older adults experience some degree of age-related vision loss, such as presbyopia or cataracts, 
and some older adults may be unable to achieve 20/20 vision even with correction due to other eye conditions such as age-related macular degeneration (AMD)~\cite{friedman2004prevalence}, it is important to balance the need to control variables with the equally-important mandate to ensure representation of older adults in the sample.

\subsection{Non-Representative Sampling Practices}
The rapid pace of innovation in visualization research may (often unconsciously) incline researchers toward \textit{samples of convenience}, which can result in the exclusion of older adults among other possible groups and populations. Inclination towards samples of convenience is a recognized pitfall in science. Sears warns psychology researchers that over-reliance on ``college sophomores in the laboratory'' can lead to a narrow and biased understanding of human nature and suggests that researchers should expand their samples to include more diverse populations and contexts~\cite{sears1986college}.
In a recent survey based on 79 visualization papers published at TVCG, CHI, EuroVis, VAST, InfoVis, VIS, and BELIV, Burns et al.~\cite{burns2023we} found that the average and median ages of participants in studies focused on novice visualization users were 30 and 28, respectively. Our intention is not to question the ethics or rigor of prior visualization studies but to stress a potential source of bias in sampling practice to which many, including the authors of this manuscript, have not been immune (e.g.,~\cite{sarvghad2015exploiting, ziemkiewicz2012visualization}).

These issues do not exist in a vacuum, but rather in the context of other forms of societal bias.
This raises an interesting dilemma, wherein attempts to broaden participation along one axis may exacerbate barriers along others. For example, consider the gradual shift from relying exclusively on in-person studies to accepting online platforms such as Amazon Mechanical Turk (mTurk) as valid environments to conduct fundamental research in VIS. 
As of 2020, only 6\% of mTurk users (minimum age 18) were aged 60 or older, compared with approximately 35.6\% of the US population~\cite{moss2020demographics}, indicating that the presence of older adults in the population of mTurk workers is far from representative. Again, our intent is not to disparage the use of these tools; many have been shown to increase access and participation by members of many groups that are historically underrepresented~\cite{bergman2016have}. 
Instead, this observation underscores the need for careful sampling in both in-person and online study populations to ensure representation across all ages.

\subsection{Influence from Stereotypes and Stigma} 
Exclusion of older adults may also arise from \textit{assumptions} about older adults' interests and abilities, as well as from \textit{stigma} about participation in research. Researchers may lean into stereotypes about older adults, such as believing they are less capable of or interested in participating in research studies~\cite{petrie2023talking,mannheim2019inclusion,levy2009stereotype}. They may also be concerned about the vulnerability of older adults and the potential risks associated with their participation, such as increased stress or discomfort~\cite{national2017nih}. 
This stereotype threat may also negatively affect older adults' perceptions of research studies as well~\cite{jacelon2007older}. 
Lack of awareness of the availability of research studies and the benefits of participation are also likely contributors, as this population is rarely the focus of direct recruiting efforts. Moreover, logistical difficulties such as access to transportation~\cite{townsley2005systematic} or venue accessibility issues also disproportionately impact older adults. Visualization researchers can use sampling methods that are more representative of the population of older adults, such as random sampling or stratified sampling that accounts for age, gender, race, and socioeconomic status. Researchers can also design studies with more inclusive eligibility criteria that allow for the participation of older adults with various health conditions and from diverse backgrounds.

\subsection{Limited Disciplinary Diversity} 
Sarvghad et al.~\cite{sarvghad2022scientometric} and Losev et al.~\cite{losev2022embracing} point to a notable lack of disciplinary diversity in visualization research. 
Although this would not necessarily directly result in the exclusion of older adults from visualization research, expanding interdisciplinary collaboration between visualization researchers and experts from different fields of study, such as gerontology and sociology, may benefit older adults' inclusion by exposing visualization researchers to aging-related research topics that intersect and complement their research interests and agendas. An interdisciplinary research approach can lead to a better understanding of the needs and preferences of older adults and promote the development of technology that is more effective and accessible~\cite{kleinman2021social}.

%% file: 3-gerontovis-vs-accessibility.tex
\section{Adjacent Areas of Study}
\label{accessibility}
\ourTerm overlaps in spirit and high-level goals with research on digital accessibility and equity.  However, its heterogeneous population and specific domain of data visualization yield a combination of requirements and considerations that likely cannot be simply covered by existing work. In this section, we discuss some more mature areas of existing research that, while not a replacement for \ourTerm, serve as an important foundation for future work.

\subsection{Graphical User Interface Design for Older Adults}
While age-inclusive data visualization research is still in its infancy, more general graphical user interface (GUI) design for older adults has a more established track record.
Some key differences between interface design and visualization have been well-documented elsewhere~\cite{liang2010exploratory, ebert2014building}, with a major source of divergence stemming from each domain's design objectives and end goals.
Graphical User Interfaces (GUIs) are primarily engineered to facilitate task completion by circumventing the need for intricate command-line instructions. In contrast, data visualizations are tailored to foster reflective thinking,  insight discovery, and data-driven decision-making~\cite{liang2010exploratory}. This difference in objective implies that design strategies suitable for older adults may not be directly transferable or equally effective across both domains.

A handful of recent works have focused specifically on interface design for older adults (e.\,g.,~\cite{mclaughlin2020designing,fisk2020designing,johnson2017designing}). 
Guidelines such as minimizing memory load, enlarging text, and careful use of color can certainly apply in a visualization context. However, in exploring these resources, a point is often reached where a design recommendation either has an unclear application to visualization or even conflicts with existing best practices.  For example, Mclaughlin and Pak recommend conveying information in multiple ways within a modality, such as color and size~\cite{mclaughlin2020designing}; this may introduce unnecessary visual complexity within the redundant data-ink~\cite{tufte2001visual}. Similarly, Czaja et al. propose reducing working memory load by displaying all pertinent information on the screen~\cite{czaja2019designing}, but the definition of ``pertinent'' can vary across visualizations, and excessive on-screen data might lead to clutter. Hence, scrutinizing the adaptability of such GUI-focused guidelines in the context of specific tasks and visualizations remains an unresolved challenge that warrants further exploration.

\subsection{Accessible Data Visualization}
Accessible data visualization is a rapidly developing field that is gaining attention from researchers and practitioners alike. Emerging areas of focus center primarily on the needs of individuals with different abilities ~\cite{marriott2021inclusive}, such as those with severe visual impairments~\cite{elmqvist2023visualization}.
\ourTerm and accessible data visualization share a common goal of promoting inclusivity and equity in data visualization. However, these two sub-disciplines must not be conflated: the lived experiences of people with disabilities (who may be of any age) and older adults (who may or may not be disabled) are not interchangeable. Older adults are a distinct population with unique characteristics, challenges, and design considerations, and we cannot assume that techniques developed with accessibility in mind will directly transfer to older adults~\cite{johnson2017designing}, nor vice versa. In particular, older adults may experience several age-related perceptual and cognitive changes simultaneously.
The progressive effects of these correlated changes can also interact and compound, leading to more complex design challenges. For example, a recent eye-tracking study of older adults observed age-associated changes in eye movement correlated with reduced manual dexterity~\cite{heintz2021role}. These observations complicate some traditional interaction paradigms, such as pinch-to-zoom: the people for whom zooming would serve as an adaptive strategy may have difficulty performing the physical gesture required to access it.

Moreover, older adults may have different preferences and expectations for how they interact with technology, and these perspectives may not be adequately represented in accessibility research with younger populations. The primary objective of \ourTerm is to create data visualizations that are not only accessible to older adults but also effective, intuitive, engaging, functional, and enjoyable. Therefore, it is important to develop interventions that account for the unique needs and preferences of older adults, rather than relying on existing accessibility solutions designed for individuals with disabilities without consideration for the effects of aging.
The benefits of investigating and designing data visualizations for older adults, however, have the potential to improve the utility of data visualization for a broad range of audiences. Well-known examples of design for specific groups that became universal design practices are curb cuts and tactile pavements~\cite{steinfeld2012universal}, as well as the adoption of web content accessibility guidelines (WCAG)~\cite{schmutz2016implementing} as part of standard best practice for the design of web-based interfaces. Understanding how to design visualizations that accommodate older adults' requirements is also a critical step in line with the call for inclusive design~\cite{clarkson2013inclusive, coleman1999inclusive}, aiming to integrate marginalized populations into mainstream design endeavors.

%% file: 4-age-related-physiological-changes.tex
\section{How Aging May Impact Visualization Use}

\label{age-related-changes}
\begin{figure}[t!]
    \ \ \ \ \includegraphics[width=0.85\columnwidth]{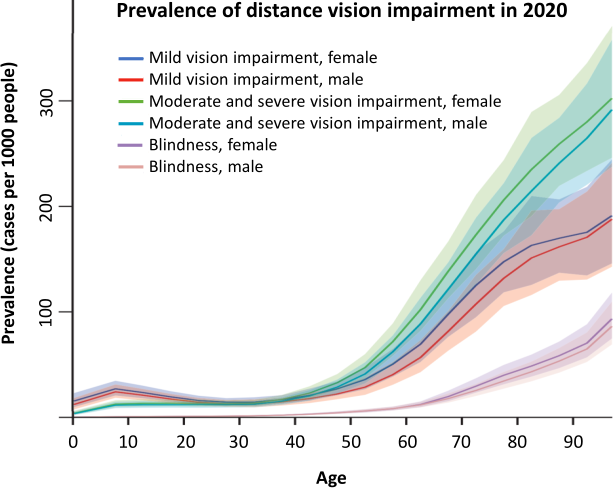}
    \caption{Global estimation of the age-specific prevalence of distance vision impairment. 
    Source:~\cite{bourne2021trends}, CC BY 4.0.}
    \label{fig:vision_impairment_trends}
\end{figure}

As we age, many people experience predictable changes in their perception, cognition, and physical capabilities. 
For example, age-related farsightedness (presbyopia) caused by the hardening of the lens and weakening of the eye's ciliary muscles is almost universal in adults 65 years and older~\cite{johnson2017designing}. \autoref{fig:vision_impairment_trends} shows the global estimation of the age-specific prevalence of distance vision impairments in 2020, illustrating a notable increase in the prevalence of moderate and severe vision impairment cases after the age of 50. 
It is important to note that the specific manifestation and rate of age-related changes vary widely among individuals and may depend on factors such as genetics, lifestyle, health status, and environmental factors. For instance, some people may not experience notable vision decline due to age.  Despite individual variations in aging, the prevalent trend suggests that most people experience such changes at a higher rate with increasing age~\cite{fisk2020designing, verhaeghen1997meta}.
In the rest of this section, we highlight how some of the age-related changes in perception, cognition, and motor function \textit{may} interact with older adults' ability to use data visualization. We would like to emphasize that the information provided in this section does not constitute a comprehensive or definitive analysis of how aging may impact older adults' use of data visualizations. Instead, it is intended to motivate \ourTerm and to illustrate the vital need for further research in order to fully investigate and understand the complex scope and nuanced intricacies of this particular subject.

\subsection{Perception}
Data visualization takes advantage of humans' high-speed perceptual processing capabilities, such as recognizing shapes, colors, and patterns to communicate data in a graphical form. Aging, however, may impact various aspects of perception in older adults. In this section, we present some of the most prevalent age-related perceptual changes and discuss how they may interfere with older adults' ability to decode and infer information from perceptual signals. 

\subsubsection{Visual Acuity} Visual acuity, or the ability to visually distinguish fine details, declines with age. This decline usually accelerates above age 50~\cite{mitzner2015considering} and can reduce the older adult's ability to discern details in visualizations. Visual acuity decline in older adults can usually be ameliorated using glasses, contact lenses, or other interventions. However, it is important to note that vision can vary among individuals, and factors such as overall eye health, refractive errors, and other age-related changes can influence visual acuity. The likelihood of co-occurrence of multiple vision issues in older adults is relatively common. Hence, even with a 20/20 corrected vision, older adults may still encounter difficulties in processing visual information, particularly for perceptually demanding tasks. For example, small-scale visualizations on compact display devices, like smartwatches, demand a high level of visual acuity to distinguish intricate graphical details, which might present difficulties for older adults with compounded age-related vision difficulties.  While this problem is not unique to visualization, it could be exacerbated by the fact that many visualizations require the user to differentiate between small, detailed elements. The decline in visual acuity may also interfere with older adults' ability to interact with visualizations, particularly on small-screen devices~\cite{mitzner2015considering}.

\subsubsection{Visual Search} Visual search refers to the process of actively scanning the visual environment to identify a specific target among distractors. 
In \textit{feature search}, a single feature identifies the target, such as looking for a red circle among green circles in a scatterplot. Generally, older adults are as fast and accurate as younger adults in performing this type of search~\cite{becic2007age}. \textit{Conjunction search}, on the other hand,  is multi-criteria, and a combination of features defines the target. Searching for a red square among red circles and green squares on a scatterplot is an example of a visualization conjunction search. This type of search is generally slower, and the time to find the target increases as the number of distractors increases for all. However, the rate of performance decline in older adults is usually more than in younger individuals~\cite{becic2007age}.  The discrepancy between younger and older adults increases for more complex triple-conjunction searches (e.g., color, shape, size)~\cite{humphrey1997age}. 

Visual search (for sighted individuals) is the fundamental perceptual activity that is the substrate for carrying out many other, more complex visualization tasks. However, we know little about what and how visualization design choices would influence visual search in older adults. For instance, both younger and older adults performing a triple-conjunction search would benefit from providing an additional dimension to define a target~\cite{humphrey1997age}. This may suggest that visualization design provisions such as double-encoding may assist older adults in more easily identifying targets. The type of task may also influence older adults' visual search performance. Older adults retain an effective ability to discern global shapes~\cite{norman2020aging}. Tasks such as identifying trends or outliers that require high-level form perception might be less demanding than tasks such as deriving values (e.g., average) that require identifying specific targets, retrieving values, and performing calculations. 

\subsubsection{Color Perception and Contrast Sensitivity} Color is among the most frequently utilized visual data encoding channels. Sensitivity to light decreases with age, particularly in the blue and green regions of the spectrum, which can make it harder for older adults to discriminate between certain hues, such as blue and purple. Long-term exposure to ultraviolet light can also cause yellowing of the lens and cornea, interfering with the perception of color, mainly yellow from white~\cite{mitzner2015considering}. Other hues could also be tinted toward yellow, making colors like green, blue, and violet hard to distinguish~\cite{johnson2017designing}. After age 50, many experience a decrease in their ability to discern subtle differences, read text, and distinguish between objects of similar hues or patterns due to a decline in contrast sensitivity. Contrast sensitivity often continues to decline, usually becoming acute by age 80~\cite{johnson2017designing}.

Color and contrast are among the more studied topics in the context of aging and data visualization (e.g.,~\cite{Reeder2014-gk, Ahmed2019-cardiac-implant, price2016effects}). However, prior work has been primarily centered around perceptual aspects of color and contrast design, overlooking questions such as the interplay between color and contrast design and the elicited emotional responses in older adults or their impact on the perceived complexity and clarity of visualizations. 

\subsubsection{Hearing}
Data sonification is a process in which data values are mapped to certain musical parameters, such as pitch, volume, and timbre~\cite{hermann2011sonification}. Sound signals can also be used as auditory cues or feedback, enabling and guiding interaction~\cite{gaver1987auditory}. Voice-enabled conversational data exploration~\cite{srinivasan2017orko}, a method of examining and analyzing data through a conversational interface, also relies on sound to communicate with the user. While audio can potentially benefit older adults with visual challenges such as those outlined in \Cref{accessibility}, common age-related auditory complications such as  hearing loss (presbycusis) may reduce the utility of sonification approaches for this population. By age 60-65, 33\% of us experience hearing loss significant enough to affect our interactions with others; this ratio jumps to 55\% for those 75-80 and 89\% for those above 80~\cite{stevens2013global}. Moreover, hearing loss is often more complex than a simple decrease in volume perception~\cite{johnson2017designing}. In presbycusis, for example, a person may have difficulty hearing certain frequencies or distinguishing sounds in noisy environments, which cannot be resolved by simply increasing the volume. Sonification for older adults might need to take this into account, using sounds within an accessible frequency range.
Additional research is needed in this area to evaluate these assumptions and to explore other facets of sonification specifically tailored for older adults. 

\subsection{Motor Control}
Age-related motor issues are most likely to impact older adults' ability to interact with visualizations. Prior research suggests that they may become frustrated or disengaged when encountering interaction issues with technology, such as difficulty using a mouse or touch screen~\cite{peek2014factors}. Interaction issues can also increase the likelihood of errors when using technology, such as accidentally clicking the wrong button or selecting the wrong option~\cite{kim2011effect}. 

\subsubsection{Manual Dexterity}
As people age, fine motor skills (e.g. those used to control a mouse or touchscreen) can decline~\cite{carmeli2003aging}. This can make it challenging for older adults to execute coordinated interactions such as multi-fingered gestures commonly used on many touchscreen devices, resulting in greater preference for single-finger gestures~\cite{stossel2010mobile}. The development of hand tremors with aging can also lead to difficulties with fine-grained touch interactions by introducing screen oscillations that make target selection increasingly hard~\cite{wacharamanotham2011evaluating}.

\subsubsection{Hand-Eye Coordination}
Older adults may experience reduced hand-eye coordination due to deteriorating vision and limb mobility with age, making it more challenging to accurately reach their interaction target, such as a specific bar in a histogram~\cite{carmeli2007function}. 
Age-related changes in eye movement characteristics, such as increased saccadic latency and decreased saccadic accuracy/velocity, may further exacerbate interaction difficulties~\cite{irving2006horizontal}.

\subsubsection{Slowing of Movement}
Older adults may experience slower reaction times when selecting specific points on a screen due to a higher number of course-corrections as they converge to a point~\cite{ketcham2002age}, which can make it more difficult to interact with visualizations that require rapid or tightly-coordinated sequential movements, such as double-clicking or dragging an element.
The influence of specific motor control changes on older adults' interactions with visualizations becomes even more critical for visualization paradigms such as ``visualization by demonstration'' ~\cite{saket2016visualization} that substitute menu-and-command-based model of visualization construction and manipulation with direct interaction methods. Health and activity monitoring visualizations on small-screen-touch-based devices, such as smartphones, are also increasingly used by older adults~\cite{moore2021older}. However, the  literature on visualization design considerations for motor decline in older adults is extremely sparse~\cite{marriott2021}.

\subsection{Cognition}
Cognitive scientists generally agree that our cognitive abilities decline with age~\cite{johnson2017designing}. However, similar to perceptual changes, the type, magnitude, and rate of cognitive changes notably vary between individuals~\cite{johnson2017designing}. Cognitive changes are also more challenging than perceptual and physical changes to detect and measure. To showcase the importance of cognitive challenges in \ourTerm, we briefly mention how the decline of working memory and attention may impact older adults' ability to use visualizations. 

\subsubsection{Working Memory} Working memory (WM) is a mechanism responsible for temporarily holding and manipulating information in our awareness. Aging can cause a decline in working memory capacity, though the starting age and the rate of decline greatly varies~\cite{salthouse2012consequences}. However, on average, the WM of older adults is lower than that of their younger counterparts~\cite{salthouse1991decomposing}. WM plays a critical role in visual information processing by allowing the brain to integrate and interpret visual stimuli over short periods of time. Lower WM capacity can interfere with recalling visual details or patterns or interpreting complex visual information~\cite{unsworth2007nature}. Age-related declines in WM can also cause difficulties in learning new information or multitasking~\cite{unsworth2007nature}, which may complicate the use of unfamiliar visualizations. Complex visualizations can be cognitively demanding, requiring users to remember and interpret multiple information pieces simultaneously. While this can be challenging for all ages, it can be particularly problematic for older individuals, who may have decreased working memory or cognitive processing speed.

\subsubsection{Attention} 
The age-related decline in various aspects of attention, such as selective and sustained attention, may also interfere with the effective use of data visualizations by older adults. As we age, our ability to filter out distractions can decline due to worsening working memory~\cite{mcnab2015age}, making it more difficult to selectively attend to relevant information in a visualization. As a result, older adults may find it more challenging to focus on the most important information in a visualization while ignoring irrelevant or distracting details such as annotations or animated transitions. Older adults may also experience declines in sustained attention due to worsening sensory function~\cite{staub2013sustained}, making it more difficult to maintain focus and concentration over an extended period of time. This can be particularly challenging for more complex visualizations that require sustained attention in order to extract meaningful insights. 

As the complexity of visualization and analysis tasks grows, the disparity in performance between older and younger adults becomes more noticeable (e.g.,~\cite{price2016effects}). This section offers an overarching discussion of how aging might influence our ability to utilize visualizations. However, many of the issues outlined here, along with others not mentioned, have yet to be empirically explored. The full scope and intricacies of age-related challenges in visualization comprehension and how to alleviate them remain poorly understood and call for further examination.

%% file: 5-state-of-the-practice.tex
\section{State of the Practice in \ourTerm}
\label{survey}

This section reviews prior work at the intersection of older adults and data visualization. Research pertaining to both older adults and data visualization is scattered across various fields of research domains. To find relevant papers, we used several research search engines: Google Scholar, IEEE Xplore, the ACM Digital Library, and the PubMed research database. While the IEEE and ACM tools provide papers within computer science, PubMed focuses on the medical and health fields, and Google Scholar is topic-agnostic with much greater scope. We included relevant search terms such as \textit{older adults}, \textit{data visualization}, \textit{aging}, \textit{seniors}, \textit{elderly}, \textit{chart}, and \textit{graph}. 
Our inclusion criteria included base necessities such as featuring or focusing on both older adults and data visualization and more specific requirements such as participants described as ``older adults'' having to be at least 60 as per the UN definition~\cite{UNHCR}. The data visualizations in these papers were either designed for and evaluated with older adults or intended for their use in specific tasks;
data visualizations \textit{about} older adults would not suffice on their own. Papers focusing on overall technology accessibility for older adults were also removed, as that area covers applications beyond data visualization. Similarly, papers purely about mobile apps and touchscreen design were removed if they made no mention of data visualizations specifically. This ultimately resulted in a set of \numPapers papers from \numVenues venues, with additional details in \Cref{tbl:domains}. The closeness of the number of papers and the number of research venues further illustrates the need for establishing \ourTerm as a uniting area of research; researchers in this area may have greater difficulty finding current literature and possible collaborators due to the spread-out nature of existing work. 

To review and report the selected publications, we first examined them to learn how data visualization was studied in relation to aging and what types of contributions were made. We observed that most research (32 out of \numPapers) was domain-specific, approaching data visualization within a particular context and specific problem. For instance, Price et al.~\cite{price2016effects} investigated how data visualization could help older adults to navigate and select between various multi-faceted Medicare plans. Perhaps due to the increase in health-related concerns that many people experience later in life, a notable portion of these works were centered around the healthcare domain. The main contributions of these works were empirical evidence (or lack thereof) of the data visualization utility for enabling older adults with problem-solving and reflections on how visualization design choices (e.g., visual encoding choice) and affordances (e.g., interactivity) would impact their performance and preference.  We also observed very few (1 out of \numPapers) domain-agnostic psychophysical experiments studying the theoretical aspects of data visualization perception and cognition in older adults. In that paper, Le et al.~\cite{le2014elementary} compared the graphical perception between younger and older adults and the general population. This paper provided empirically-driven knowledge of visualization design for older adults, such as performance-based ranking.

\begin{table}[t!]
\centering
\caption{This table provides information about the set of \numPapers papers we found that focus on older adults and data visualization.}
\vspace{-0.5em}
\label{tbl:domains}
\begin{tabular}{p{.15\textwidth}p{.08\textwidth}p{.08\textwidth}
} 
\hline
\Centering{\textbf{Domain}} & \Centering{\textbf{\# Papers}} & \Centering{\textbf{\# Venues}} \\
\hline
Health Informatics & \Centering{14} & \Centering{11} \\
Computer Science & \Centering{11} & \Centering{8} \\
Psychology & \Centering{6} & \Centering{6} \\
Medicine & \Centering{5} & \Centering{4} \\
\hline
\end{tabular}
\vspace{-1.5em}
\end{table}

\subsection{Visual Encoding}
Visual encoding is the process of mapping data values to various properties of graphical marks. Visual encoding channels such as position, color, angle, and text serve as the building blocks for creating meaningful visual representations of data.
In a study with 32 older adults, Price et al.~\cite{price2016effects} concluded that using color to encode important information could reduce working memory load, thus facilitating the decision-making process and improving decision quality. However, Alexander et al.~\cite{alexander-passive} observed that older adults had difficulty working with color maps with similar hues and low-contrast designs. 
Similarly, the legibility of axis labels, annotations, and other text is an important consideration when designing visualizations for older adults. 
Reading smaller fonts grows progressively difficult with age~\cite{johnson2017} and several studies point to older adults' challenges using visualizations due to this issue~\cite{fan2023understanding, morey-heart-failure, whitlock}. Enlarging font size is an immediate solution for legibility issues, but may not always be feasible due to constraints such as limited screen real state (e.g., smartwatch). A possible approach for visualization designers and practitioners to meet older adults' font requirements is to link visualizations with existing device accessibility features or offer other types of assistance, such as magnification or screen readers. 

Previous investigations of color and text provide initial insights into the topic of visual encoding within the context of visualization design for older adults. However, several crucial questions regarding the visual encoding process and the selection of visual encoding channels still remain unexplored. Recent research by Davis et al.~\cite{davis2022risks} challenges the universal applicability of existing rankings of visual encoding channels, as they observed significant variations in individuals' performance on visualization tasks. Older adults are a population characterized by their heterogeneity, making it essential to understand how the choices of visual encoding and their effectiveness correlate with different characteristics and requirements within this demographic. Further exploration is needed to uncover these relationships and enhance our understanding of effective visual encoding strategies for older adults.

\subsection{Contrast Ratio and Contrast Polarity}
Established guidelines (e.g.,~\cite{wcag2018}) offer recommendations for designing digital information interfaces for older adults. For instance,  Web Content Accessibility Guidelines (WCAG)~\cite{wcag2018} recommends a minimum of 4.5:1 contrast ratio for text and 3:1 for non-text elements when designing web content for older adults. However, data visualizations targeted at this demographic do not consistently follow these guidelines~\cite{fan2023understanding}.

Morey et al.~\cite{morey-heart-failure} evaluated the accessibility of two popular mobile health apps, learning that both apps had low color contrast that hindered their usability for older adults. Whitlock and McLaughlin~\cite{whitlock} discovered similar contrast and color choice issues during a hierarchical task analysis on three mobile health (mHealth) apps, recommending a change of dark-grey-on-light-grey text and lines to black-on-white based on existing accessibility guidelines for older adults~\cite{fisk2004,pak2010}. Fan et al.~\cite{fan2023understanding} also identified low contrast as a design issue that hindered older adults' ability to gain insights from online COVID-19 visualizations. In a set of semi-structured interviews conducted with retirement home residents, Reeder et al.~\cite{Reeder2014-gk} collaboratively designed a set of smart home data visualizations. They cited Whitlock and McLaughlin when choosing black and white for maximum contrast during the visualization design process~\cite{whitlock}, additionally finding that participants struggled when relying solely on color to differentiate thin bars. Focus groups run by Le et al.~\cite{le-2015-eval} had difficulty interpreting values from gradient color maps. 

Contrast polarity refers to the level of contrast difference between foreground objects and the background, with positive contrast featuring light objects on a dark background (known as dark mode) and negative contrast featuring dark objects on a light background (known as light mode). Previous research in the field of vision science has examined the impact of contrast polarity choices on various aspects of older adults' performance, such as reading speed and accuracy (e.g.,~\cite{chung2009contrast, piepenbrock2013positive}), ability to group disjointed
contour segments belonging to a single object~\cite{spehar2002role}, and ability to read in older adults with low vision~\cite{rubin1989psychophysics}. Empirical evidence from these studies suggests that negative contrast is generally a more favorable choice for older adults. However, there is currently a lack of comparable research evaluating the effects of contrast polarity design choices in data visualizations on the performance and preference of older adults. 

\subsection{Visualization Complexity and Visual Aids}
Several empirical studies (e.\,g.,~\cite{jones-2012,whitlock, morey-heart-failure}) advocate for the adoption of simple and high data-ink ratio visualizations as a means to alleviate cognitive load and demand on older adults.
In a participatory design project conducted by Cajamarca et al.~\cite{cajamarca2023understanding}, older adults expressed their preference for health data apps on smartwatches to have ``familiar,'' ``clear,'' and ``clean'' designs. In two separate studies, Le et al.~\cite{le-2015-eval, le-2018-understanding} also found that older adults tended to focus on aggregate measures and data overviews unless trying to explain an observed pattern or change, indicating that older adults' information-seeking behavior may conform to Schneiderman's visualization mantra~\textit{``overview first, details on demand''}~\cite{shneiderman2003eyes}. Less complex and familiar visualizations can reduce short-term memory demand for comparisons and lower cognitive load in older adults with declining working memory~\cite{fabiani2012, Le2012-integrated, le2014elementary, tao2018, wu2021}. However, older adults' inclination towards streamlined visualizations is also attributed to factors not related to their cognitive functions, such as visualization literacy~\cite{hakone2016}. In addition to using simpler visualization types, prior work also suggests reducing unnecessary embellishments, such as removing boxes around text~\cite{Reeder2014-gk, fan2023understanding}.

While prior literature advocates for using simpler visualizations and removing unnecessary embellishments, there are also arguments toward augmenting visualizations with additional elements and cues such as annotations and interaction signifiers (e.g., ~\cite{tao2018, le-2018-understanding, le-2015-eval, fan2022accessibility}).
These provisions could help older adults to understand data better or work with interactive visualizations. For instance, Tao et al.~\cite{tao2018} found that including text annotations led to older adults' improved understanding of data. In a paper outlining guidelines for data visualizations for older adults, Le et al.~\cite{le-2015-eval} also recommend using visual aids to bring attention to especially-noteworthy data points. Finally, Huh et al.~\cite{Huh2013-rp} and Alexander et al.~\cite{alexander-passive} suggest including guidance for older adults with low computer and visualization literacy in order to help them discover tools and visualizations' interactive functions.
Balancing the tension between reducing clutter and adding scaffolding is an interesting open challenge in the context of visualization design for older adults. 

There is also no consensus regarding what constitutes a ``familiar'' visual encoding. While bar, line, and pie charts have been shown to be perceived as familiar and easy-to-understand by older adults~\cite{doyle-2015, Reeder2014-gk, Ahmed2019-cardiac-implant}, other visualizations based on familiar concepts, such as circles with different brightness and size to present data, were met with more skepticism~\cite{le-2015-eval}. We hypothesize that familiarity is more heavily informed by lived experience than by chronological age (though, of course, these are not independent). For example, older adults in a participatory study by Ahmed et al.~\cite{Ahmed2019-cardiac-implant} used a stoplight metaphor (red, yellow, and green) to visually encode healthy, average, and unhealthy values, stating that familiarity with the semantics of traffic lights would help them to quickly and accurately interpret the information. However, participants in this same study also articulated a preference for viewing temporal data using a bar chart instead of the more ubiquitous line chart. In addition to counter-intuitive \textit{preferences}, the literature also contains contradictory evidence regarding \textit{performance} using familiar metaphors~\cite{le2014elementary,poirier}.

Lacking contextual information may hinder older adults' ability to interpret perceived information. For instance, in the context of health and wellness visualizations, older adult participants across several independent studies reported uncertainty regarding what was the expected or ``good'' data range~\cite{Ahmed2019-cardiac-implant,alexander-passive, fan2023understanding}. Participants also report uncertainty when trying to contextualize their data in comparison to others~\cite{Reeder2014-gk} and lamented a lack of context for observed changes in their own health data~\cite{le-2018-understanding}. In addition to external context for individual values, a study of older adults presented with heart health data surfaced a desire to understand better the relationships \textit{between} various data dimensions as well. The initial guidelines presented by Le et al.~\cite{le-2015-eval} recommended offering methods of data comparison and clearly depicting changes over time. Doyle et al.~\cite{doyle-2015} noted that enabling older adults to annotate their own charts and input their own data can lead to better understanding. 

\subsection{Sense-Making, Communication, and Decision-Making}
Several of the prior work studies the value and utility of data visualization for supporting patient-provider communication, medical decision-making, and self-health monitoring and managing various facets of their health and wellness, such as chronic heart issues~\cite{Ahmed2019-cardiac-implant, morey-heart-failure}.
Researchers have enumerated some of the perceived benefits and barriers to the adaptation and use of data visualization tools for health monitoring and management from the perspective of older adults and healthcare providers. Older adults found these tools helpful for communicating with their physician, caretakers, and family~\cite{reeder2013, Huh2013-rp} and noticing otherwise-undetected and temporal changes in their health~\cite{doyle-2015, le-2018-understanding}. Among the primary factors hindering older adults'  use and perceived usefulness of these tools are low visualization and computer literacy~\cite{Huh2013-rp}, low confidence in learning and using new technology~\cite{Huh2013-rp, bock2016}, lack of context for interpreting data~\cite{doyle-2015, le-2018-understanding}, privacy concerns~\cite{Huh2013-rp}, and lack of transparency of how tools use their data~\cite{Huh2013-rp}. 

Visualizations have been used to support communication between older adults and their healthcare providers, primary caregivers, and family members.
In a study with groups of five older adults, family members, nurses, and doctors, Alexander et al.~\cite{alexander-passive} observed that family members appreciated the ability to use a web-enabled data visualization dashboard to check their resident's health and physical activity remotely, and the older adults felt safer with the knowledge that someone else was keeping track of any worrisome changes. Both groups agreed that communicating the data would benefit medical caretakers, and all participants were willing to share the data with relevant stakeholders. Huh et al.~\cite{Huh2013-rp} also found that both older adults and medical professionals agreed that visual communication of personal health data could strengthen the rapport between patients and physicians by allowing healthcare providers to gain a deeper insight into a patient's current condition and medical history. These studies also surfaced some concerns and challenges related to collecting and communicating personal health data, such as lack of context to interpret the data by older adults and their family members~\cite{alexander-passive}, concerns over privacy~\cite{le-2018-understanding}, and increased anxiety due to overexposure to their health data~\cite{doyle-2015}. Huh et al.~\cite{Huh2013-rp} also emphasized the importance of creating tools with different user interfaces designed for different stakeholder groups (patient, family, nurse, doctor) to make these tools much more robust and valuable.

In a set of focus groups with older adults, Le et al.~\cite{le-2015-eval} observed participants' positive responses and inclination toward using a wellness dashboard of data visualizations to assist them in collaborative decision-making with their physicians.
Price et al.~\cite{price2016effects} observed that visualizing information about different Medicare plans improved older adults' ability to compare alternatives and select a plan that best matched their requirements.
They posit that by reducing the working memory load with a visualization, older adults used a more systematic approach to making the decision instead of employing simple heuristics. They also observed that when making more ``high difficulty'' decisions, providing visual-based assistance was not enough to significantly increase accuracy for older adults, concluding that solely reducing cognitive load may not be enough in these cases.
Van Weert et al.~\cite{van2021preference} also recommend augmenting visualizations with other data communication methods, such as tables, to increase motivation and interest for  those with lower visualization literacy.

In addition to investigating various use cases of data visualization in health care, these also offer valuable insights that can inform the design of visualizations for older adults. For instance, Ahmed et al.~~\cite{Ahmed2019-cardiac-implant} observed that older adults consistently removed a sad-looking face emoji from a heart health dashboard due to its negative connotation and concluded that such designs may discourage continued use, while Cajamarca et al.\cite{cajamarca2023understanding} observed that older adults preferred health visualizations that instilled tranquil feelings. While studying how older adults would use data visualizations for self-assessment of their abilities to use a computer mouse, Jones et al.~\cite{jones-2012} observed that participants noticeably found charts depicting hypothetical unwanted health scenarios to be unfavorable and avoided interacting with them.
Further research is necessary to learn design considerations that would promote a positive outlook toward visualization use by older adults~\cite{hawleyhauge2014}. 

The outcomes of the \numPapers publications reviewed in this section provide valuable insight into the design  and utility of data visualizations for older adults. Nevertheless, a majority of these works concentrate on discrete and individual facets of aging and their influence on older adults' utilization of data visualization. This focus overlooks the complex reality that people often experience multiple age-related changes and challenges simultaneously. Consequently, the cumulative and compounded effects of aging extend far beyond isolated issues. Older adults are also a heterogeneous population, and the type and efficacy of interventions may vary substantially between different sub-populations~\cite{petrie2023talking}. Hence, A one-size-fits-all research approach will likely fail to meet the varying needs and requirements of this demographic.

%% file: 6-challenges-and-opportunities.tex
\section{Challenges and Opportunities}
\label{challenges-and-opportunities}

In \Cref{age-related-changes}, we discussed how some of the age-related changes in perception, cognition, and physical abilities might interfere with the older adults' ability to use data visualizations.
The existing breadth of research into these phenomena's biological triggers and general effects is well understood (e.g., \cite{madden2004age, mitzner2015considering, unsworth2007nature}). However, the way in which compounded effects of these changes influence older adults' ability to read and understand visualizations, particularly as the complexity of visualization and tasks increases, is not always clear. There is also a dearth of empirically driven and tested knowledge of designing functional and effective data visualizations and interactions for older adults. Using existing standards and guidelines for designing digital content (e.g.,~\cite{wcag2018, ibm-carbon-design}), general digital technology (e.g.,~\cite{wagner2010computer, williams2013considerations, nunes2012design}), and interactive visual interfaces (e.g.,~\cite{mclaughlin2020designing}) as a guide, evaluating the utility and adaptation of such recommendations in the context of visualization design for older adults is another avenue for narrowing our \ourTerm knowledge gap.

Acknowledging the existence and importance of these knowledge gaps is the first necessary step toward bridging them. One possible strategy to expedite \ourTerm' research is replication. Replication studies will enable us to re-examine the validity, transferability, and limitations of our existing visualization knowledge, assumptions, and guidelines for older adults. They can also help surface new challenges related to the ability and requirements of older adults within the context of the replicated studies. While a plausible starting point, replication alone is not enough for the multifaceted growth and maturity of \ourTerm. Replication studies are also associated with known difficulties, such as
obtaining the necessary data and information about the original study design. Hence, such efforts should be accompanied by studies designed and carried out with older adults as their primary focus. 

Characterizing and measuring visualization literacy for older adults is another open knowledge challenge. The topic of visualization literacy has been gaining increasing attention from the community (e.g.,~\cite{solen2022scoping}), with ongoing debates and discussions regarding how to define and measure it. Current visualization literacy assessment platforms such as VLAT~\cite{lee2016vlat} assume a uniform underlying capability of the users to execute the required tasks; in other words, they assume that the provided visualizations are eminently usable for all participants. However, it is unlikely that a one-size-fits-all approach to measuring visualization literacy would adequately accommodate the diverse population of older adults. Reexamining existing visualization literacy and examination techniques with older adults can be a first step in this direction.

\subsection{Methodological Challenges}
Conducting research with older adults requires specific methodological considerations due to their unique attributes, needs, and abilities~\cite {quine2007methodological, quinn2010methodological, levy2003conceptual}. Hence, a critical component of successfully integrating older adults into visualization research is learning and employing necessary methodological adjustments and considerations that meet and accommodate their characteristics and requirements. Such considerations may pertain to various aspects of visualization research design and activities, including sampling and recruitment, study design, and evaluation techniques and metrics. For instance, visualization researchers should be cognizant of and sensitive to the inherent diversity of the older adult population in their sampling approaches. As people age, they encounter various life experiences and health conditions that can affect their physical, cognitive, and social functioning differently. Older adults are generally considered to be a more heterogeneous population than younger adults~\cite{whitbourne2010adult, rowe2015successful}. While representative samples are universally important, \ourTerm researchers should specifically pay close attention to this matter. Techniques such as purposeful or quota sampling that can increase the diversity and inclusion of older adults with varying profiles should be part of the routine \ourTerm practice. 
The design of visualization studies should also consider older adults' requirements and characteristics. For instance, between-subjects studies are not usually recommended for behavioral studies due to the heterogeneity of the older adult population~\cite{dickinson2007methods}. Think-aloud data collection protocol can also cause problems for adults who struggle with multitasking~\cite{johnson2017designing}. 

Using larger fonts, high-contrast colors, and simplified language can make questions and instructions easier to follow and reduce the chance of interference with participants' study tasks. Older adults may also require more time and support for performing tasks, particularly if they have physical or cognitive impairments. Controlled lab experiments may cause anxiety and stress in older adults, especially when instructions are unclear and confusing~\cite{le-2016-eval}. Therefore, the data collection process may need to be more flexible and adaptable to their needs. Measures and metrics such as time and accuracy that are commonly used to assess the performance of younger adults may not accurately capture that of older adults.

Incorporating human-centered methods, such as focus groups and participatory design workshops, can enhance the effectiveness of visualization design for older adults. However, existing participatory methodologies are primarily tailored towards younger adults~\cite{gregor2001designing, czaja2019designing}, and thus may not be suitable for engaging older adults. Developing new methodologies and approaches that cater to older adults' specific needs and limitations may be necessary for successful participatory visualization research with this population. For instance, older adults may face mobility and transportation challenges that hinder their ability to participate in design sessions outside their homes. To overcome this barrier, a flipped model that brings studies to older adults may be a more feasible and effective approach. Tools designed to aid contextual studies, such as the \textit{lab-in-a-box}~\cite{weibel2015lab}, which can serve as a model for developing tools that facilitate on-site visualization research with older adults.
To this end, we advocate for the inclusion of \ourTerm into ongoing conversations on evaluation and methodology in visualization, such as IEEE BELIV~\cite{BELIV20294:online}.

\subsection{Ethical Challenges}
Ethical considerations and moral codes of conduct are other areas of interest to \ourTerm. Performing data visualization studies with older adults raises ethical questions and concerns related to topics such as informed consent, privacy and confidentiality, locus of control, and stigma and discrimination. Special care must be taken in obtaining informed consent from older adults who are experiencing age-related changes in cognitive function, visual acuity, or auditory processing, as this may affect their ability to fully understand the research and its implications~\cite{sugarman1998getting}. 

This population may also be more vulnerable to power imbalances in research if they depend on others for care~\cite{ray2007redressing}. This challenge requires developing mechanisms and protocols to ensure visualization research does not (inadvertently) exploit or disadvantage older adults and to promote respectful and equitable interactions with them. 
It is also important to avoid ``parachute research'' that prioritizes publishing papers on a target population instead of meaningfully involving and benefiting them~\cite{lundgard2019sociotechnical}.
Research in this area should not just enrich our knowledge of visualization design for older adults but be translated into real-world solutions that benefit and empower older adults in practical day-to-day needs.

Previous work has also shown that the privacy of personal data is a major concern of older adults, particularly for healthcare visualizations that could possibly be shared with family members, stakeholders, and medical professionals~\cite{Huh2013-rp, le-2018-understanding, alexander-passive}. Older adults have showcased strong interest and care about the handling of their own data \cite{wang2019}, so it is important for both researchers and practitioners to provide options and give careful consideration to how user data for older adults is kept when running experiments and designing tools for this population. 
We must establish frameworks and guidelines for \ourTerm researchers to navigate these complex issues and help them to avoid behavioral pitfalls such as (unconsciously) perpetuating stigma and discrimination based on ageism and negative stereotypes about aging.

\subsection{The Gap Between \ourTerm Research and Practice}
A recent survey by Fan et al.~\cite{fan2023understanding} found that the use of colors that were hard to distinguish for aging vision (e.g., blue-green), low contrast ratio between graphical elements, and small font size were naming the reason that hindered older adults' ability to utilize and gain insights from COVID-19 visualizations effectively. This is despite the existence of visualization design guidelines and recommendations specific to the above-mentioned issues and research studies that have highlighted the importance of accommodating older adults' visual and cognitive needs. Lim et al.~\cite{lim2023establish} work also emphasizes a crucial gap between research and practice, ``particularly in the sphere of aging''. To enable older adults to fully benefit from data visualization, it is not only necessary to close the existing gap in our knowledge but also to investigate and understand factors that will encourage and result in practical adaptation and use of such knowledge in the real world.

While laboratory experiments provide controlled conditions for studying how older adults interact with data visualization, it is equally crucial to investigate their encounters and usage in natural settings, such as their daily living environments. Older adults increasingly encounter graphical displays of information in a variety of contexts, including health monitoring and self-management~\cite{cajamarca2020}, patient-provider communication~\cite{reeder2013}, medical decision-making~\cite{price2016effects}, smart home self-monitoring~\cite{le2014design}, news articles~\cite{fan2023understanding}, and home devices~\cite{baskinger2007experientializing}. The prevalence of digital displays makes it essential to understand how older adults navigate increasingly ubiquitous small screens, complex interfaces, and low contrast in daily life. This "in the wild" approach to research underscores the important of investigating complex user behavior in the contexts and environments in which that behavior manifests.

%% file: 7-conclusion.tex
\section{Conclusion}
In this paper, we introduced~\textit{\ourTerm}, an area of visualization research and practice focusing on the inclusion and empowerment of older adults. Designing data visualizations for older adults calls for attention to understanding how aging can influence their ability to use visualizations and what considerations would improve the utility and usefulness of visualization for this demographic. We discussed how some age-related physiological changes in perception, cognition, and motor control might hinder older adults' ability to use visualizations effectively. We also provide a high-level review of prior work at the intersection of aging and data visualization, identify some topics and challenges investigated, and discuss their implications for designing visualizations for older adults. Our discussion of challenges and opportunities offers a starting point for dialogue and future research in this area. 

In discussing the physiological and cognitive changes associated with aging that may influence how older adults use visualizations, we have also reflected on some of the potential sources of the systematic exclusion of older adults in visualization research, including (but not limited to) bias based on those physiological changes. Our intention in this work is to challenge the stereotypes that portray older adults as less capable or disinterested, particularly with respect to their interactions with visualization and digital technology in general. Recognizing that older adults are an engaged, vibrant, and dynamic population, we call for increased attention to this demographic in visualization research. 
 
 We conclude this work with the simple observation that while the specific trajectory may vary widely for each individual, aging is a near-\textbf{universal facet of the human experience}. As we age, it is perhaps also a universal hope that the major structural components of our lives will have comparable usability in old age as they did when we were younger. By investigating the ways age-related physiological changes affect how we use and understand data visualizations, we are not only addressing the needs of the older adults present in our communities now; we are also setting up our future selves for success. Establishing and building upon \ourTerm makes important strides toward a more equitable future where data visualization is \textit{truly} for everyone.